\documentclass[aps,prd,superscriptaddress,nofootinbib,amsmath,amsfonts,preprintnumbers,notitlepage,10pt,english]{revtex4-1}
\usepackage{amssymb}
\usepackage{graphics}
\usepackage{graphicx}
\usepackage{amsmath}
\usepackage{amsfonts}
\usepackage{bm}% bold math

\newlength{\extraspace}
\newlength{\extraspaces}
\newcommand{\be}{\begin{equation}
\addtolength{\abovedisplayskip}{\extraspaces}
\addtolength{\belowdisplayskip}{\extraspaces}
\addtolength{\abovedisplayshortskip}{\extraspace}
\addtolength{\belowdisplayshortskip}{\extraspace}}
\newcommand{\ee}{\end{equation}}
\newcommand{\ba}{\begin{eqnarray}
\addtolength{\abovedisplayskip}{\extraspaces}
\addtolength{\belowdisplayskip}{\extraspaces}
\addtolength{\abovedisplayshortskip}{\extraspace}
\addtolength{\belowdisplayshortskip}{\extraspace}}
\newcommand{\ea}{\end{eqnarray}}
\usepackage[figtopcap]{subfigure}

\usepackage{color,soul}
\usepackage[colorlinks,citecolor=blue,urlcolor=blue,linkcolor=blue]{hyperref}
\newcommand{\nonu}{\nonumber \\[.5mm]}

\usepackage{scalerel}
\usepackage{tikz}
\usetikzlibrary{svg.path}
\definecolor{orcidlogocol}{HTML}{A6CE39}
\tikzset{
  orcidlogo/.pic={
    \fill[orcidlogocol] svg{M256,128c0,70.7-57.3,128-128,128C57.3,256,0,198.7,0,128C0,57.3,57.3,0,128,0C198.7,0,256,57.3,256,128z};
    \fill[white] svg{M86.3,186.2H70.9V79.1h15.4v48.4V186.2z}
                 svg{M108.9,79.1h41.6c39.6,0,57,28.3,57,53.6c0,27.5-21.5,53.6-56.8,53.6h-41.8V79.1z M124.3,172.4h24.5c34.9,0,42.9-26.5,42.9-39.7c0-21.5-13.7-39.7-43.7-39.7h-23.7V172.4z}
                 svg{M88.7,56.8c0,5.5-4.5,10.1-10.1,10.1c-5.6,0-10.1-4.6-10.1-10.1c0-5.6,4.5-10.1,10.1-10.1C84.2,46.7,88.7,51.3,88.7,56.8z};}}
\newcommand\orcid[1]{\href{https://orcid.org/#1}{\mbox{\scalerel*{
\begin{tikzpicture}[yscale=-1,transform shape]
\pic{orcidlogo};
\end{tikzpicture}
}{|}}}}
\begin{document}
\date{\today}

\title{The general expression for $f(T)$ in a charged cylindrical spacetime with diverse dimensions}

\author{G.~G.~L.~Nashed~\orcid{0000-0001-5544-1119}}
\affiliation{Centre for theoretical physics, the British University in Egypt, 11837 - P.O. Box 43, Egypt}
% Center for Space Research, North-West University, Potchefstroom 2520, South Africa}
\hspace{2cm} \hspace{2cm}
\begin{abstract}
By utilizing the field equations of the modified teleparallel equivalent of general relativity, denoted as $\mathit{f(T)}$, we obtain an exact solution for a static charged black hole in n-dimensions, without imposing any constraints.  The black hole possesses two distinctive dimensional constants: $m$ and $v$ with unit {\textit length}. The first constant is associated with the mass, while the second constant represents the electric charge. The existence of this electrical charge causes the black hole to diverge from the expectations of the teleparallel equivalent of general relativity (TEGR). Our analysis demonstrates  that $\mathit{f(T)}$ is reliant on the parameter $v$ and transforms into a constant expression when $v$ is assigned a value of zero. A captivating aspect of this particular black hole is its absence of singularities in the quantities formed using torsion and curvature, given that the dimension $n$ falls within the interval of $4 \leq n \leq 6$ as $r$ approaches zero.  However, for $n\geq7$, the singularity becomes milder in comparison to the case of TEGR. Furthermore, By utilizing the conserved n-momentum vector, we calculate the energy of this solution and confirm its correspondence with the ADM mass, accurate to the order of $O\Big(\frac{1}{r}\Big)$.  Otherwise, we observe higher-order contributions arising from the electric charge terms. Through the application of a coordinate transformation to the black hole, we derive a precise solution describing a stationary rotating black hole. This solution showcases significant readings of the torsion scalar and the analytical function $\mathit{f(T)}$. In order to gain insight into the physics of this black hole, we calculate various physical quantities related to thermodynamics, such as entropy, Hawking temperature, and heat capacity. The analysis reveals that the black hole exhibits thermal stability.
\end{abstract}

\maketitle

%%%%%%%%%%%%%%%%%%%%%%%%%%%%%%%%%%% Section 1 %%%%%%%%%%%%%%%%%%%%%%%%%%%%%%%%%%%%%%%%
\section{Introduction}\label{S1}
%%%%%%%%%%%%%%%%%%%%%%%%%%%%%%%%%%%%%%%%%%%%%%%%%%%%%%%%%%%%%%%%%%%%%%%%%%%%%%%%%%%%%%

%arXiv:1806.11445v2#
General relativity (GR) theory, that depends on the curvature, has shown great successful in the realm of astrophysics as well as in cosmology \cite{Abbott:2016blz,Abbott:2017oio,TheLIGOScientific:2017qsa,Weinberg:1972kfs,1968ApJ...151..431M}. It is well know that there is another theory, completely equivalent to GR from the viewpoint of the field equation, which is called  teleparallel equivalent
of general relativity (TEGR) \cite{Einstein:1918btx,article,Einstein1930}. In the domain of cosmology,  both  GR and TEGR have  shown defects to describe the history of the universe in a consistent way \cite{Riess:1998cb,Perlmutter:1998np,Ade:2013zuv,Spergel:2006hy,Jain:2003tba,Cole:2005sx,Eisenstein:2005su,2010MNRAS.401.2148P,Padmanabhan:2012hf,2011MNRAS.418.1707B,2013MNRAS.428.1036M,PhysRevD.71.123001,Stern:2009ep,Zhang:2012mp,Blake:2011ep,Chuang:2012qt,Moresco:2012jh,2003MNRAS.346...78H,PhysRevD.69.103501,Cole:2005sx}. Therefore, it becomes necessary to modify  GR or TEGR theory to resolve the issues that  both of them  could not explain.

Many modifications have been  constructed in order to face the  previous defects. Among these modifications that constructed in the frame of Riemannian geometry  is  $\mathit{f(R)}$ gravitational theory which has many successful applications in  cosmology and astrophysics \cite{1980PhLB...91...99S,Nojiri:2003ft,Nojiri:2003ni,Capozziello:2003gx,Capozziello:2002rd,Carroll:2003wy,PhysRevD.77.046009,PhysRevD.77.023503,PhysRevD.73.123504,PhysRevD.75.084010,PhysRevD.76.063517,PhysRevD.78.043002,Nojiri:2007cq,Nojiri:2007as,PhysRevLett.101.061103,PhysRevD.80.064002,PhysRevD.82.064033,PhysRevD.92.064015,Bamba:2008ut,Capozziello:2006dj,PhysRevD.93.023501,Capozziello:2012ie,Capozziello:2011et,Nojiri:2010wj,Nojiri:2006ri,Nojiri:2017ncd}. There is another modification in the context of Weitzenb\"{o}ck geometry which is coined as $f(T)$ gravitational theory \cite{PhysRevD.79.124019}. There are many successful cosmological  applications using $f(T)$ theory, { with $T$ being the torsion scalar} \cite{Cai:2015emx,Linder:2010py,Chen:2010va,
2012GReGr..44.3059M,2012GReGr..44.3059M,Zheng:2010am,Bamba:2010wb,Addazi:2023pfx,
Dent:2011zz,Cai:2011tc,Li:2011rn,Wu:2011kh, Wei:2011aa,Liu:2012fk,Capozziello:2017bxm, Amoros:2013nxa,
Otalora:2013dsa,Li:2013xea,2014EPJP..129..188N,Jusufi:2022loj,Ong:2013qja,Nashed:2014lva,Kofinas:2014daa,Darabi:2014dla,
Nashed:2015pda, Haro:2014wha, DeBenedictis:2018wkp,Nashed:2018efg,Capozziello:2023vne,
Guo:2015qbt,Malekjani:2016mtm,Farrugia:2016qqe,Nashed:2003ee,
Qi:2017xzl,Sk:2017ucb,
 Bahamonde:2017wwk, Karpathopoulos:2017arc,
Hohmann:2017jao,Hohmann:2018rwf,Cai:2018rzd,Gonzalez:2011dr,Capozziello:2012zj,Iorio:2012cm,Nashed:2013bfa,Aftergood:2014wla,
Paliathanasis:2014iva,2015IJMPD..2450007N,Capozziello:2022zzh,
2015EPJP..130..124N,
Junior:2015fya,Nashed:2016tbj,Awad:2017sau,Ahmed:2016cuy,Farrugia:2016xcw,Mai:2017riq,
Awad:2017tyz,Bejarano:2017akj,Arcos:2004tzt}. The $f(T)$ gravitational theory has the ability to yield dark energy and dark matter  \cite{Bengochea:2008gz,2010PhRvD..81l7301L,Wu:2010av,2011JCAP...01..009D,2011JCAP...01..021B,2010arXiv1008.4036B,2012PhRvD..85j4036B,Aviles:2013nga,Jamil:2012vb,Nashed:2018cth,Ferraro:2011us,Ferraro:2011zb,Sebastiani:2010kv,
 Salako:2013gka,Haghani:2012bt,Haghani:2013pea}. The equations governing the field in the context of the $f(T)$ theory, as deliberated in \cite{Bengochea:2008gz, Linder:2010py, Cai:2015emx}, are second-order equations, in contrast to the fourth-order equations found in $\mathit{f(R)}$ theories \cite{Nojiri:2003ft, Nojiri:2003ni}. This is noteworthy since the $f(T)$ theory can accommodate an arbitrary torsion scalar term.  In spite of the fact that there is a correspondence
between GR and TEGR on the field equations, their
modifications $\mathit{f(R)}$ and $\mathit{f(T)}$ are not equal \cite{Cai:2015emx}. { Analytical solutions of the field equations in $f\left( T\right) $-gravity within a cosmological context have been obtained using the approach of movable singularities in differential equations. The examined models include $f_{1}\left( T\right) =T+\alpha\left( -T\right) ^{n}$ and $f_{2}\left( T\right) =T+\alpha\left( -T\right) ^{n}-\Lambda$, with the retrieval of GR  achieved as $\alpha$ approaches zero \cite{Paliathanasis:2016vsw}. An investigation has been conducted to demonstrate the presence and enduring nature of the Kasner vacuum solution of Bianchi type I in the context of modified $f\left( T\right) $ teleparallel gravity \cite{Paliathanasis:2017htk}. A thorough introduction is provided on the development of teleparallel geometry as a gauge theory of translations, along with an exploration of all the other characteristics of gauge field theory \cite{Bahamonde:2021gfp}.}

In the early years of the last century, the theory of extra dimensions was developed and formulated in \cite{Kaluza:1921tu, Klein:1926tv}. To the best of our current understanding, there exists no precise mathematical representation for $\mathit{f(T)}$ that characterizes a charged cylindrical spacetime. This inquiry seeks to attain a comprehensive equation for the function $\mathit{f(T)}$ across spacetimes of varying dimensions, with a specific focus on cylindrical spacetime defined by two unconstrained functions.   The structure of this study is as follows: In Section \ref{S2}, the fundamental framework of $\mathit{f(T)}$ gravitational theory is presented. Section \ref{S3} utilizes a vielbein (tetrads) with flat transverse sections to apply it to the n-dimensional charged field equations of $\mathit{f(T)}$. Consequently, an analytic solution for a static n-dimensional charged black hole is derived, which exhibits asymptotic behavior similar to that of de Sitter/anti-de Sitter spacetime. This black hole possesses an intriguing characteristic wherein a constant of integration establishes a connection between the higher-order terms of the electric charge. When this constant becomes zero, the black hole reverts to the black hole found in TEGR theory. In Section \ref{S4}, our attention turns to examining the significant physical properties of the  solution obtained in Section \ref{S3}. This is accomplished by studying the singularities of the scalars  constructed from torsion and curvature.  What stands out prominently is that within the range of $4\leq n\leq 6$, the invariants formed from curvature and torsion present no instances of singularity, while for $n\geq 7$, the singularity manifests with reduced severity when contrasted with the TEGR scenario.
 %In  Section \ref{S5}, we apply a coordinate transformation to the black hole and derived analytic rotating black hole that has non-trivial values of torsion and $\mathit{f(T)}$.
 In Section \ref{S6}, we explore the thermodynamic properties of the black hole introduced in Section \ref{S3} and demonstrate its thermal stability. The last section concludes this research with final remarks.
%%%%%%%%%%%%%%%%%%%%%%%%%%%%%%%%%%% Section 2 %%%%%%%%%%%%%%%%%%%%%%%%%%%%%%%%%%%%%%%%
\section{Basics of $f(T)$  gravitational theories}\label{S2}
%%%%%%%%%%%%%%%%%%%%%%%%%%%%%%%%%%%%%%%%%%%%%%%%%%%%%%%%%%%%%%%%%%%%%%%%%%%%%%%%%%%%%%
The Lagrangian for the gravitational theory $\mathit{f(T)}$ is provided as follows\footnote{ In this study we will not include in the Lagrangian of $ \mathit{ f(T)}$ gravitational theory the Lagrangian of the matter, ${\cal L}_{_{_ Matter}}$, because we are interested in the vacuum case.} \cite{Bengochea:2008gz,2012PhRvD..85j4036B}
\begin{equation}\label{q7}
\mathit{\cal L}=\mathit {\frac{1}{2\kappa}\int |b|(f(T)-2\Lambda)~d^{n}x+\int |b|{\cal L}_{ em}~d^{n}x},
\end{equation}
with  $\kappa$  being a constant which has the form \begin{align}\mathit{\kappa =2(n-3)\Omega_{n-1} G_n}\,, \quad {\mbox where} \quad  \mathit {G_n},\end{align}  is the   constant of gravitational field and $\mathit {\Omega_{n-1}}$ represents the volume of  $(n-1)$-dimensional, that has the form \begin{align}\mathit {\Omega_{n-1} = \frac{2\pi^{(n-1)/2}}{\Gamma((n-1)/2)}}.\end{align}  In Eq. (\ref{q7}) $ \mathit {|b|=\sqrt{-g}=\det\left({b^i}_\nu\right)}$ and $\mathit {{\cal L}_{
em}=-\frac{1}{2}{ F}\wedge ^{\star}{F}}$ is the  Lagrangian of Maxwell field,
where \[\mathit {F = dA}, \qquad {\mbox with}  \qquad \mathit {A=A_{\alpha}dx^\alpha},\] being the electromagnetic
potential vector \cite{Capozziello:2012zj}. Variation of Eq. (\ref{q7})  regarding the vielbein field $\mathit{{b^i}_\nu}$ gives the field equations \cite{Bengochea:2008gz}
\begin{eqnarray}\label{q8}
& &\mathit  {E^\nu_\mu={S_\mu}^{\rho \nu} \partial_{\rho} T f_{TT}+\left[b^{-1}{b^i}_\mu\partial_\rho\left(b{b_i}^\alpha
{S_\alpha}^{\rho \nu}\right)-{T^\alpha}_{\lambda \mu}{S_\alpha}^{\nu \lambda}\right]f_T
-\frac{f-2\Lambda}{4}\delta^\nu_\mu+4\pi{{{\cal
T}^{{}^{{}^{^{}{\!\!\!\!\scriptstyle{em}}}}}}}^\nu_\mu,}\nonumber\\
&& \mathit  {\equiv  f_T G^\nu_\mu+\frac{1}{2} \delta^\nu_\mu [f-f_T T]-f_{TT}{S^\nu{}_\mu}^{\alpha}\nabla_\alpha T+4\pi{{{\cal
T}^{{}^{{}^{^{}{\!\!\!\!\scriptstyle{em}}}}}}}^\nu_\mu,}\nonumber\\
&&L^\mu\equiv\mathit  {\partial_\nu \left( \sqrt{-g} F^{\mu \nu} \right)=0,}
\end{eqnarray}
where $ \mathit {f := f(T)}$, \ \   $ \mathit {f_{T}:=\frac{\partial f(T)}{\partial T}}$, \ \  $ \mathit{f_{TT}:=\frac{\partial^2 f(T)}{\partial T^2}}$ and $ \mathit{{{{\cal
T}^{{}^{{}^{^{}{\!\!\!\!\scriptstyle{em}}}}}}}^\nu_\mu}$ is the 
energy-momentum tensor of the  electromagnetic field. The supperpotential tensor $ \mathit{{S_\mu}^{\rho \nu}}$ is defined as \begin{align}\mathit{{{S_\alpha}^{\mu \nu} := \frac{1}{2}\left({K^{\mu\nu}}_\alpha+\delta^\mu_\alpha{T^{\beta
\nu}}_\beta-\delta^\nu_\alpha{T^{\beta \mu}}_\beta\right)}}.\end{align}  The contorsion, denoted as $\mathit{{K^{\mu\nu}}_\alpha}$, is defined as  \begin{align} \mathit{{K^{\mu \nu}}_\alpha  :=
-\frac{1}{2}\left({T^{\mu \nu}}_\alpha-{T^{\nu
\mu}}_\alpha-{T_\alpha}^{\mu \nu}\right)},\end{align}  and $\mathit {{T^{\mu \nu}}_\alpha}$ is the torsion tensor that has the form  \begin{align}\mathit{{T^\alpha}_{\mu \nu} =
\overset{\mathbf{w}}{\Gamma}{}^\alpha{}_{\nu\mu}-\overset{\mathbf{w}}{\Gamma}{}^\alpha{}_{\mu \nu} ={b_i}^\alpha
\left(\partial_\mu{b^i}_\nu-\partial_\nu{b^i}_\mu\right)}\,, \qquad {\mbox where} \qquad \mathit{\overset{\mathbf{w}}{\Gamma}{}^\alpha{}_{\nu\mu}},\end{align} is the curvatureless
Weitzenb{\"{o}}ck connection.

  Equation  (\ref{q8}) can be rewritten   as
\be  \mathit {\partial_\nu \Biggl\{e{S}_{a \rho}{}^ \nu f_T\Biggr\}=\kappa b
{e_a}^\mu \Biggl\{t_{\rho \mu}+{{{\cal
T}^{{}^{{}^{^{}{\!\!\!\!\scriptstyle{em}}}}}}}_{\rho \mu}\Biggr\},}\ee
where $\mathit{t_{\nu \mu}}$ has the form \be\mathit{ t_{\nu
\mu}=\frac{1}{\kappa}\Biggl[4f_T {S^\alpha}{}_\nu
{}^\lambda{T_{\alpha \lambda}}{}_{\mu}-g_{\nu \mu} f\Biggr].}\ee Due to the fact that the tensor  $\mathit{{S}_{a \nu \lambda}}$ is skew-symmetric tensor in the last two indices,   therefore \be \label{q1}
\mathit{\partial^\mu \partial^\nu\left[e{S}_{a \mu \nu} f_T\right]=0,} \quad
 \textrm{that \quad gives}
\quad \mathit{\partial^\nu\left[b\left(t^a{}_ \nu+{{{\cal
T}^{{}^{{}^{^{}{\!\!\!\!\scriptstyle{em}}}}}}}^a
{}_\nu\right)\right]=0.} \ee Equation (\ref{q1}) yields \be \label{q2}
\mathit{\frac{d}{dt}\int_V d^{(n-1)}x \ b \ {b_a}^\mu \left(t^0{}_ \mu+{{{\cal
T}^{{}^{{}^{^{}{\!\!\!\!\scriptstyle{em}}}}}}}^0{}_
\mu\right)+ \oint_\Sigma \left[b \ {b_a}^\mu \ \left(t^j{}_
\mu+{{{\cal T}^{{}^{{}^{^{}{\!\!\!\!\scriptstyle{em}}}}}}}^j{}
\mu\right)\right]=0.}\ee Equation (\ref{q2}) represents the conservation principle for the energy-momentum tensor $\mathit{{{{\cal
T}^{{}^{{}^{^{}{\!\!\!\!\scriptstyle{em}}}}}}}^{\nu
\mu}}$  and   the quantity $\mathit{t^{\lambda \mu}}$. Therefor,  one can think of the term $\mathit{t^{\nu \mu}}$
 to prescribes  the energy-momentum tensor of the field of gravity in the frame of
  $\mathit{f(T)}$  gravitational theory \cite{Ulhoa:2013gca,Nashed:2004pn,Nashed:2007sc,Nashed:2008ys}.  Thus,
the   energy-momentum tensor of $\mathit{f(T)}$ gravitational theory involved
in   (n-1)-dimensional volume $V$  has the form \be \label{q3} \mathit{P^a=\int_V d^{(n-1)}x
\ b \ {b^a}_\nu \left(t^{0 \nu}+{{{\cal
T}^{{}^{{}^{^{}{\!\!\!\!\scriptstyle{em}}}}}}}^{0
\nu}\right)=\frac{1}{\kappa}\int_V d^{(n-1)}x  \partial^\nu\left[b{S}^{a 0}{}_
\nu f_T\right].}\ee By setting $\mathit{f(T) = T}$ in equation (\ref{q3}), one can recover the TEGR case, as discussed in \cite{Maluf:2002zc, Nashed:2020kjh}. In this investigation, we will be examining the vacuum scenario within  $\mathit{f(T)}$ gravitational theory.

%%%%%%%%%%%%%%%%%%%%%%%%%%%%%%%%%%% Section 3 %%%%%%%%%%%%%%%%%%%%%%%%%%%%%%%%%%%%%%%%
\section{Cylindrical charged solutions in $\mathit{f(T)}$ theory}\label{S3}
%%%%%%%%%%%%%%%%%%%%%%%%%%%%%%%%%%%%%%%%%%%%%%%%%%%%%%%%%%%%%%%%%%%%%%%%%%%%%%%%%%%%%%

 Utilizing the field equations with charge from the $\mathit{f(T)}$ gravitational theory, represented by Eq. (\ref{q8}), to the cylindrical spacetime with n dimensions, we derive the vielbein within cylindrical coordinates ($t$, $r$, $\psi_1$, $\psi_2$, $\cdots$, ${ \psi_{n-2}}$) as outlined below \cite{Capozziello:2012zj}:
\begin{equation}\label{tetrad}
\hspace{-0.3cm}\begin{tabular}{l}
  $\left({b_{i}}^{\mu}\right)=\left( \sqrt{a(r)}, \; \frac{1}{\sqrt{a_1(r)}}, \; r, \; r, \; r\;\cdots \right)$
\end{tabular}
\end{equation}
with $\mathit{a(r)}$ and $\mathit{a_1(r)}$ are two unknown functions of $r$. The structure of the metric for the vielbein (\ref{tetrad}) can be expressed as follows:
 \be \label{met1}\mathit{ ds{}^2=a(r)dt^2-\frac{1}{a_1(r)} dr^2-r^2{\sum_{i=1}^{n-2}}d\psi^2_i}.\ee
 By employing Eqs. (\ref{tetrad}), we compute the torsion scalar as\footnote{For brevity we will use $\mathit{a(r)}\equiv \mathit{a}$,  \ \ $\mathit{a_1(r)}\equiv \mathit{a_1}$, \ \ $\mathit{a'\equiv\frac{da}{dr}}$ and $\mathit{a'_1\equiv\frac{da_1}{dr}}$ .}
\begin{equation}\label{df}
{\mathrm T(r)}=(n-2)\frac{a'a_1}{ra}+(n-2)(n-3)\frac{a_1}{r^2}.
\end{equation}
Using Eq. (\ref{tetrad}) in Eq. (\ref{q8})   we obtain the following  non-vanishing components of the charged field equations of $f(T)$ as:
%\newpage
\begin{eqnarray}\label{df1}
& &\mathit{ E_{t}{}^t=\frac{2(n-2)a_1f_{TT} T'}{r}-\frac{(n-2)f_T}{r^2a}\Biggl\{2(n-3)aa_1+ra_1a'+raa'_1\Biggr\}+f-2\Lambda-\frac{2a_1q'^2}{a}=0,}\nonumber\\
& &\mathit{E_{r}{}^r= 2Tf_T+2\Lambda-f+\frac{2a_1q'^2}{a}=0,}\nonumber\\
& & \mathit{E_{\psi}{}^\psi=I_{\psi_1}{}^{\psi_1}\cdots =E_{\psi_{n-2}}{}^{\psi_{n-2}}= \frac{f_{TT} [r^2T+(n-2)(n-3)a_1]T'}{(n-2)r}}\nonumber\\
& & \mathit{+\frac{f_T}{2r^2{a}^2}\Biggl\{2r^2aa_1a''-r^2a_1a'^2+2(2{n}-5)raa_1a'+r^2aa'a'_1+2(n-3)a^2[2(n-3)a_1+ra'_1]\Biggr\}-f+2\Lambda-
\frac{2a_1q'^2}{a}=0\,,}\nonumber\\
%
%& &{L^{t}=-\frac {r a'_1  a q' -ra_1  a'q' +2\,ra_1a  q''+4\,a_1  a\,q' }{2r a^{2}}=0\,,}
\end{eqnarray}
 where $q$ is an unknown function of the redial coordinate $r$ that is constructed from  the ansatz of the vector potential as:
\begin{equation} A=q(r)dt\,.
\end{equation}
%Equations (11) are four non-linear differential equations with four unknown functions, $a$, $a_1$, $q$, and $f(T)$. So, there must be an exact unique solution for this system.
{  To derive the general solution of Eq. (\ref{df1})  we are going to employ the following chain rule:\footnote{ Because we are working in cylindrical spacetime, the unknown functions are  depend only on the radial coordinate, $r$, and thus the torsion scalar, $T$.}
 %As a result, we will employ the fact that $f(T(r))$ is dependent on the radial coordinate, i.e., $f(r)$. }}
\begin{eqnarray}\label{dfg}
&&\mathit{f(T)=f_1(r)},\nonumber\\
& & \mathit{f_T=\frac{df(T)}{dT}=\frac{df_1(r)}{dr}\frac{d r}{dT}=\frac{r^3 a^2 f_1'}{(n-2)[r^2aa'a'_1+(n-3)ra^2a'_1+r^2aa_1a''-raa_1a'-2(n-3)a^2a_1-r^2a_1a'^2]},}\nonumber\\
& & \mathit{f_{TT}=\frac{df_T}{dT}=\frac{d}{dr}\Bigg(\frac{df_1(r)}{dr}\frac{d r}{dT}\Bigg)=\frac{r^5 a^3}{(n-2)^2[r^2a_1[aa''-a'^2]+raa'[ra'_1-a_1]+(n-3)a^2[ra'_1-2a_1]]^{^{^3}}}\times} \nonumber\\
& &\mathit{\times\Big[ r a f_1''\Big(r^2aa_1a''-r^2a_1a'^2+raa'\{ra'_1-a_1\}+(n-3)a^2\{ra'_1-2a_1\}\Big)-r^3f_1'a^2a_1a'''+f_1'\Big(3r^3aa_1a'}\nonumber\\
& &\mathit{-2r^2a^2a''[ra'_1-a_1]-r^2a^2a''_1[ra'+(n-3)a]-2r^3a_1a'^3
+2r^2aa'^2[ra'_1-a_1]+2ra^2a'[ra'_1-a_1]}\nonumber\\
& &\mathit{+2(n-3)a^3[2ra'_1-3a_1]\Big)\Big] ,}\nonumber\\
& &
\end{eqnarray}}
{ where $\mathit{f_1'=\frac{df_1(r)}{dr}}$ and  $\mathit{f_1''=\frac{d^2f_1(r)}{dr^2}}$. If we use Eq. (\ref{dfg}) in Eq. (\ref{df1}),  we get the following resulting non-linear differential equations:

\begin{eqnarray}\label{df11}
& &\mathit{ E_{t}{}^t=\frac{2(n-2)a_1}{4r^4\left( a_1  {r}^{2}a a''  -a_1   a'^{2}{r}^{ 2}+ra   \left( ra'_1 -a_1   \right) a' + \left( ra'_1  -2a_1  \right)  a^{2} \right)^{3}}\left[{r}^{5} \left\{ f_1'{r}^{3} a^{2}a_1  a'''  -r \left\{ a_1  {r}^{2}a a''  -a_1 a'^{2}{r}^{2}\right.\right.\right.}\nonumber\\
&&\mathit{\left.\left.\left.+ra   \left( ra'_1  -a_1  \right) a'  + \left(  ra'_1 -2a_1   \right)  a^{2} \right\} a  f_1''  + f_1'  \left[  \left\{ 2{r}^{2} a^{2} \left(ra'_1  -a_1   \right)-3 {r}^{3} a' a_1  a   \right\}a''  +{r}^{2} a^{2} \left(  a'  r+a   \right) a'_1  +2{r}^{3} a'^{3}a_1\right.\right.\right.}\nonumber\\
&&\mathit{\left.\left.\left.  -2{r}^{2}a   \left(ra'_1 -a_1   \right)  a'^{2}-2r a^{2} \left( ra'_1  -a_1  \right) a'  -2 \left( 2ra'_1   -3a_1  \right)  a^{3} \right]  \right\} a^{3}\right]\left[2a_1  {r}^{2}a  a'  -2a_1 a'^{2}{r}^{2}+2ra   \left(r a'_1 -a_1    \right) a'\right. }\nonumber\\
&&\mathit{\left. + 2 \left( ra'_1  -2a_1 \right) a^{2}{ r}^{3} a^{2}\right]+\frac{(n-2) f_1'  {r}a^{2}\left\{2(n-3)aa_1+ra_1a'+raa'_1\right\}}{a\left(2a'_1{r}^{2}a a'  + a'_1 r a^{2}+a_1  {r}^{2}a  a''  -a_1  ra  a' -2a_1 a^{2}-a_1  a'^{2}{r}^{2}\right)}-f_1+2\Lambda+\frac{2a_1q'^2}{a}=0,}\nonumber\\
& &\mathit{E_{r}{}^r=\frac{4a_1 a f_1'  {r}a^{2}\left( a'r+a  \right) }{\left(2a'_1{r}^{2}a a'  + a'_1 r a^{2}+a_1  {r}^{2}a  a''  -a_1  ra  a' -2a_1 a^{2}-a_1  a'^{2}{r}^{2}\right)}+2\Lambda-f_1+\frac{2a_1q'^2}{a}=0,}\nonumber\\
& & \mathit{E_{\psi}{}^\psi=I_{\psi_1}{}^{\psi_1}\cdots =E_{\psi_{n-2}}{}^{\psi_{n-2}}= \frac{ 2a_1  \left( a'r+a  \right) +a(n-2)(n-3)a_1}{a(n-2)r}\left[{r}^{5} \left\{ f_1'{r}^{3} a^{2}a_1  a'''  -r \left\{ a_1  {r}^{2}a a''  -a_1 a'^{2}{r}^{2}\right.\right.\right.}\nonumber\\
&&\mathit{\left.\left.\left.+ra   \left( ra'_1  -a_1  \right) a'  + \left(  ra'_1 -2a_1   \right)  a^{2} \right\} a  f_1''  + f_1'  \left[  \left\{ 2{r}^{2} a^{2} \left(ra'_1  -a_1   \right) -3 {r}^{3} a' a_1  a   \right\}a''  +{r}^{2} a^{2} \left(  a'  r+a   \right) a'_1  +2{r}^{3} a'^{3}a_1\right.\right.\right.}\nonumber\\
&&\mathit{\left.\left.\left.  -2{r}^{2}a   \left(ra'_1 -a_1   \right)  a'^{2}-2r a^{2} \left( ra'_1  -a_1  \right) a'  -2 \left( 2ra'_1   -3a_1  \right)  a^{3} \right]  \right\} a^{3}\right]\left[2a_1  {r}^{2}a  a'  -2a_1 a'^{2}{r}^{2}+2ra   \left(r a'_1 -a_1    \right) a'\right. }\nonumber\\
&&\mathit{\left. + 2 \left( ra'_1  -2a_1 \right) a^{2}{ r}^{3} a^{2}\right]}\nonumber\\
& & \mathit{-\frac { f_1'  {r}}{2(2a'_1{r}^{2}a a'  + a'_1 r a^{2}+a_1  {r}^{2}a  a''  -a_1  ra  a' -2a_1 a^{2}-a_1  a'^{2}{r}^{2})}\Biggl\{2r^2aa_1a''-r^2a_1a'^2+2(2{n}-5)raa_1a'}\nonumber\\
& & \mathit{+r^2aa'a'_1+2(n-3)a^2[2(n-3)a_1+ra'_1]\Biggr\}+f_1-2\Lambda+\frac{2a_1q'^2}{a}=0\,,}\nonumber\\
&&\mbox {and the differential equation of the electric field yields the form:}\nonumber\\
&&{L^{t}=-\frac {r a'_1  a q' -ra_1  a'q' +2\,ra_1
 a  q''+4\,a_1  a\,q' }{2r a^{2}}=0\,.}
\end{eqnarray}
The system of differential equation (\ref{df11}) are four differential equations in four  unknown $a$, $a_1$, $q$, and $f(r)$ which means that there is an exact  unique solution that has the form\footnote{The solution of the system of differential equations (\ref{df11}) has been checked by the software GRTensor.}:}
\begin{eqnarray} \label{sol}
 &&\mathit{a(r)=r^2c_1 -\frac{c_2{}^{n-3}}{(n-1)r^{n-3}}-\frac{c_3{}^{2(n-3)}}{2(n-2)r^{2(n-3)}}-\frac{c_3{}^{3n-8}}{3(n-2)r^{3n-8}}-\frac{c_3{}^{3(n-2)}}{(3n-4)r^{3(n-2)}}-
 \frac{c_3{}^{3n-4}}{(3n-2)r^{3n-4}},} \nonumber\\
& &\mathit{ a_1(r)=\frac{1}{\Big[1+\frac{(2n-5)c_3{}^{n-2}}{(n-3)r^{n-2}}+\frac{(2n-3)c_3{}^{n}}{(n-3)r^{n}}+\frac{(2n-1)c_3{}^{n+2}}{(n-3)r^{n+2}}\Big]^2}\Big(r^2c_1 -\frac{c_2{}^{n-3}}{(n-1)r^{n-3}}-\frac{c_3{}^{2(n-3)}}{2(n-2)r^{2(n-3)}}-\frac{c_3{}^{3n-8}}{3(n-2)r^{3n-8}}}\nonumber\\
& &\mathit{-\frac{c_3{}^{3(n-2)}}{(3n-4)r^{3(n-2)}}-
 \frac{c_3{}^{3n-4}}{(3n-2)r^{3n-4}}\Big)\equiv \frac{a(r)}{\Bigg[1+\frac{(2n-5)c_3{}^{n-2}}{(n-3)r^{n-2}}+\frac{(2n-3)c_3{}^{n}}{(n-3)r^{n}}+\frac{(2n-1)c_3{}^{n+2}}{(n-3)r^{n+2}}\Bigg]^2},}\nonumber\\
 && {\mathit{q(r)}=\frac{c_3{}^{n-3}}{r^{n-3}}+\frac{c_3{}^{2n-5}}{r^{2n-5}}+\frac{c_3{}^{2n-3}}{r^{2n-3}}+\frac{c_3{}^{2n-1}}{r^{2n-1}},}\nonumber\\
& & \mathit{f(r)=2\Lambda-\frac{2\Big[\frac{(n-3)c_3{}^{2(n-2)}}{r^{2(n-2)}}+\frac{(2n-5)c_3{}^{3(n-2)}}{3r^{3(n-2)}}+\frac{(n-4)(2n-3)c_3{}^{(3n-4)}}{(3n-4)r^{(3n-4)}}
+\frac{(n-6)(2n-1)c_3{}^{(3n-2)}}{(3n-2)r^{(3n-2)}}\Big]}{(n-3)\Bigg[1+\frac{(2n-5)c_3{}^{n-2}}{(n-3)r^{n-2}}+\frac{(2n-3)c_3{}^{n}}{(n-3)r^{n}}+\frac{(2n-1)c_3{}^{n+2}}{(n-3)r^{n+2}}\Bigg]},}
\end{eqnarray}
{ where $\mathit{c_1}$, $\mathit{c_2}$, and $\mathit{c_3}$ are   dimensionalfull constants of unite ${\textit length}$. The vector potential { $q(r)$} shows that the parameter $c_3$ represents the electric charge. When the constant $c_3$ is equal zero we recover the uncharged black hole presented in \cite{Nashed:2019zmy}.}

%%%%%%%%%%%%%%%%%%
\section{The physical implications of the  solution given in equation (\ref{sol})}\label{S4}
\begin{figure}
\centering
\includegraphics[scale=0.4]{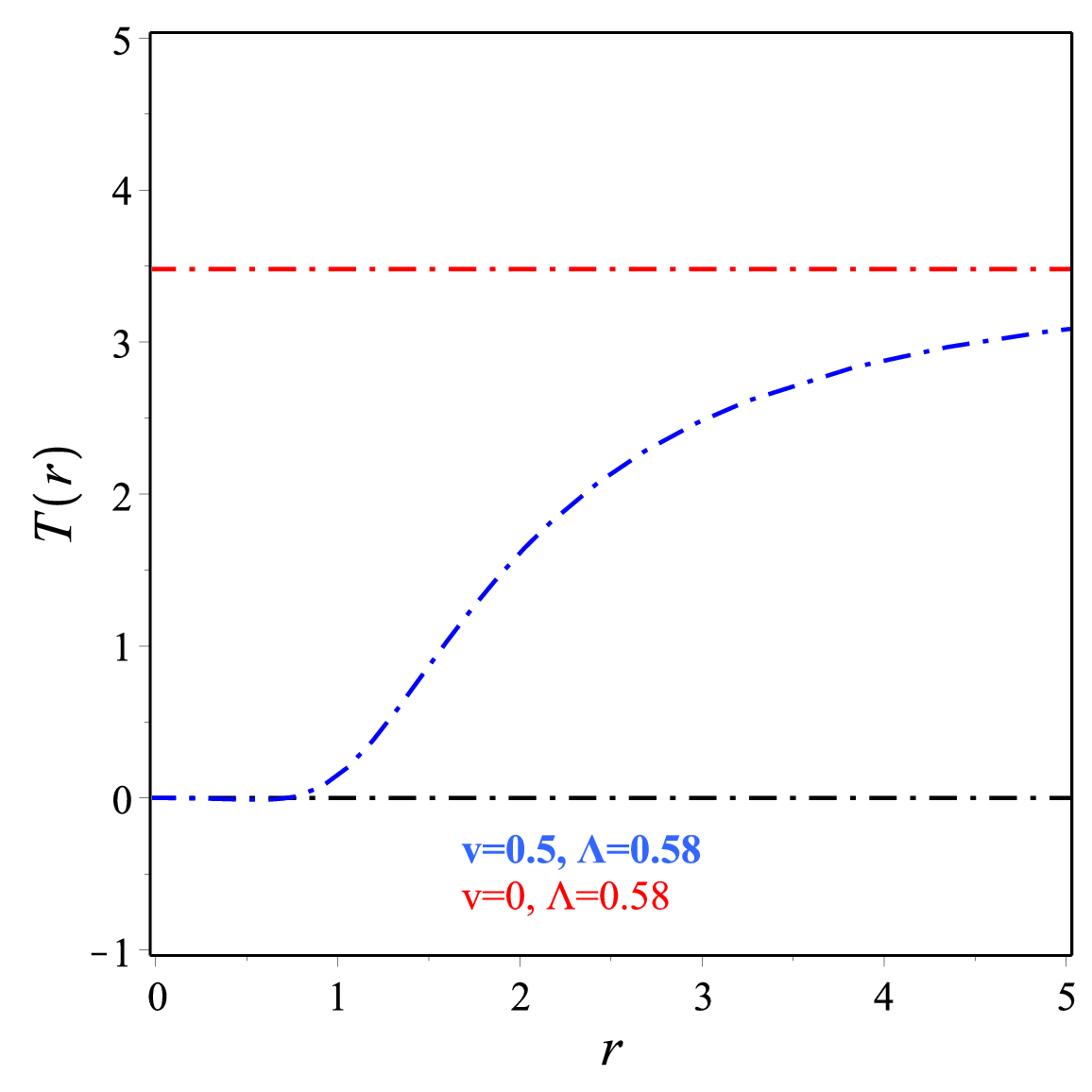}
%\subfigure[~The spherically symmetric AdS/dS spacetime]{\label{fig:1b}\includegraphics[scale=0.4]{JMTh2}}
\caption{Schematic plot of the torsion scalar of the black hole (\ref{sol})}
\label{Fig:1}
\end{figure}

\begin{figure}
\centering
\subfigure[~The analytic function $f(r)$ given by (\ref{sol})]{\label{fig:2a}\includegraphics[scale=0.35]{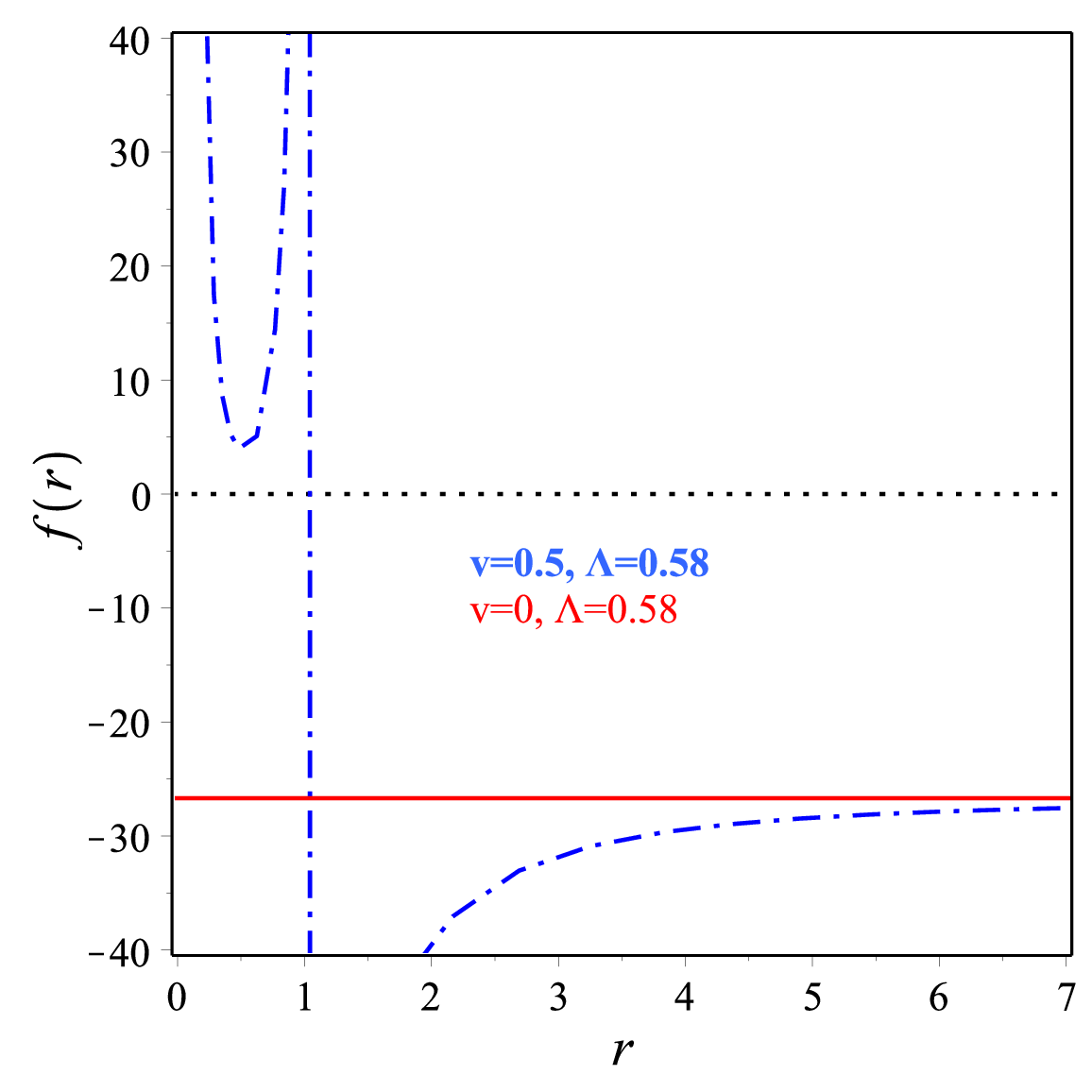}}
\subfigure[~The analytic function $f(T)$ given by (\ref{fT})]{\label{fig:2b}\includegraphics[scale=0.35]{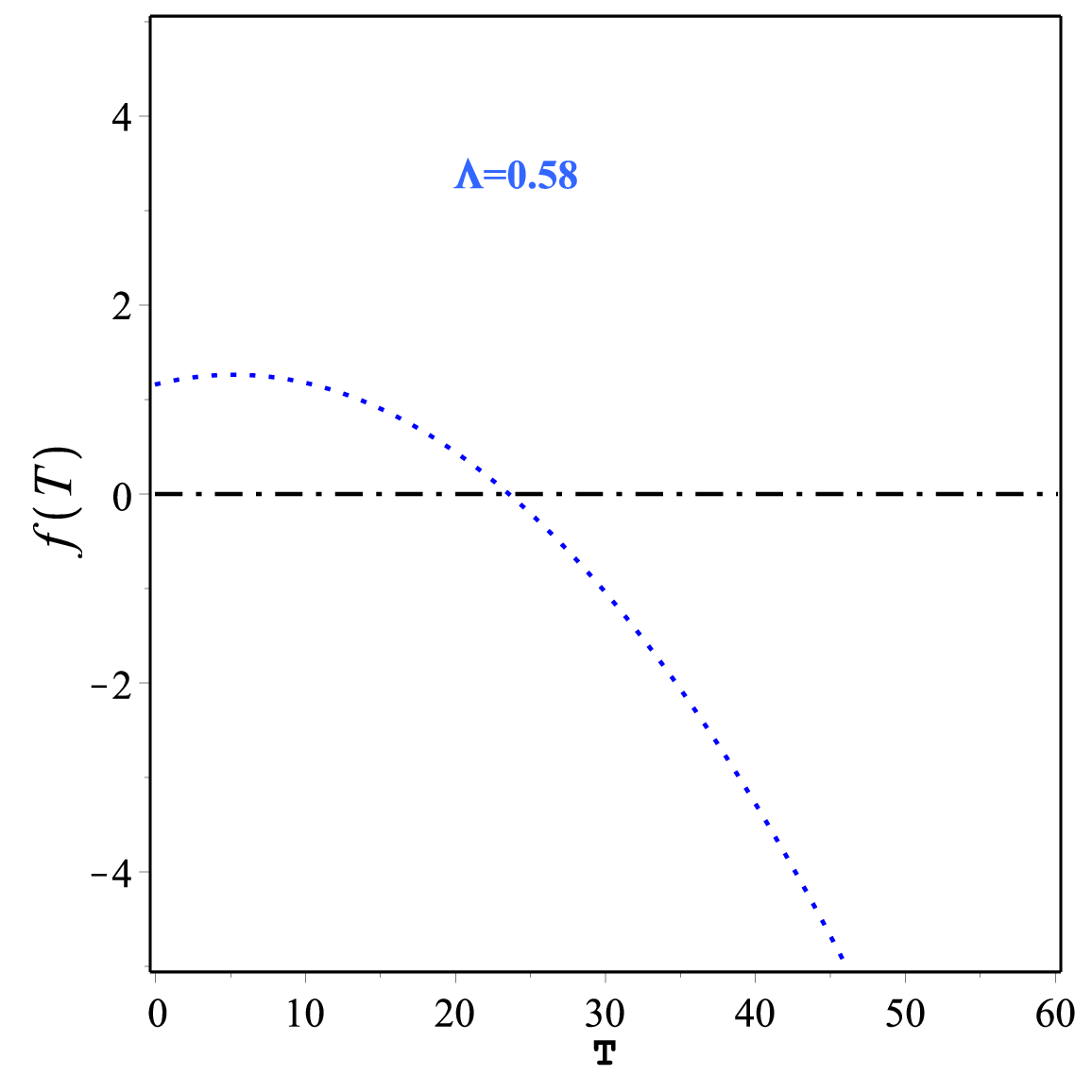}}
\caption{Schematic plot of the functions $f(r)$ and $f(T)$.}
\label{Fig:2}
\end{figure}

Within this section, we will delve into an examination of the pertinent physics associated with the solution acquired in the preceding section. \vspace{0.2cm}\\
\underline{Scalar torsion:}\vspace{0.2cm}\\
From  Eqs. (\ref{df}) and (\ref{sol}) the scalar torsion yields:
\be \label{Tor1} T(r)=\frac{(n-1)(n-2)\Lambda+\frac{(n-3)c_3{}^{2({n}-2)}}{2r^{2({n}-2)}}+\frac{(2n-5)c_3{}^{3({n}-2)}}{3r^{3({n}-2)}}+\frac{(2n-3)(n-2)c_3{}^{(3n-4)}}{(3n-4)r^{(3n-4)}}
+\frac{(2n-1)(n-2)c_3{}^{(3n-2)}}{(3n-2)r^{(3n-2)}}}{(n-3)^2\Big[1+\frac{(2n-5)c_3{}^{{n}-2}}{(n-3)r^{{n}-2}}
+\frac{(2n-3)c_3{}^{{n}}}{(n-3)r^{{n}}}+\frac{(2n-1)c_3{}^{({n}+2)}}{(n-3)r^{({n}+2)}}\Big]^{^2}}.\ee
Equation (\ref{Tor1}) indicates that the torsion scalar can remain constant when {$c_3=0$}, resulting in the vanishing contribution of the electric field in that scenario. The behavior of the torsion scalar is depict in figure \ref{Fig:1}.  It is imperative to announce that  it is not easy to solve Eq. (\ref{Tor1}) to find $r$ in terms of the torsion scalar $T$. However, in the 4-dimension case, we can find the asymptotic behavior of Eq. (\ref{Tor1})  up to $O(\frac{1}{r^4})$ and then substitute in Eq. (\ref{sol}) and find the corresponding $f(T)$ which is plot  in  figure \ref{Fig:2}\subref{fig:2b}.   \vspace{0.2cm}\\
\newpage
\underline{The metric spacetime:}\vspace{0.2cm}\\
The metric of the vielbein  (\ref{tetrad}), after using Eqs. (\ref{sol})  in Eq. (\ref{met1}), yields
\begin{eqnarray}  \label{met2}
 && \mathit{ds{}^2=\Biggl(r^2\Lambda -\frac{m^{n-3}}{r^{n-3}}+\frac{v^{2(n-3)}}{2(n-2)r^{2(n-3)}}+\frac{v{}^{3n-8}}{3(n-2)r^{3n-8}}+\frac{v{}^{3(n-2)}}{(3n-4)r^{3(n-2)}}+
 \frac{v{}^{3n-4}}{(3n-2)r^{3n-4}}\Biggr)dt^2}\nonumber\\
&&\mathit{-\Bigg[\frac{1}{\Big\{1+\frac{(2n-5)v^{n-2}}{(n-3)r^{n-2}}+\frac{(2n-3)v{}^{n}}{(n-3)r^{n}}+\frac{(2n-1)v{}^{n+2}}{(n-3)r^{n+2}}\Big\}^2}\Big(r^2\Lambda -\frac{m^{n-3}}{r^{n-3}}+\frac{v^{2(n-3)}}{2(n-2)r^{2(n-3)}}+\frac{v{}^{3n-8}}{3(n-2)r^{3n-8}}}\nonumber\\
& &\mathit{+\frac{v{}^{3(n-2)}}{(3n-4)r^{3(n-2)}}+
 \frac{v{}^{3n-4}}{(3n-2)r^{3n-4}}\Big)\Bigg]^{-1}dr^2-r^2\sum_{i=1}^{{n}-2}d\psi^2_i,} \end{eqnarray} where we have put $\mathit{c_1=\Lambda}$, $\mathit{c_2=m(n-1)}$,  and  $\mathit{c_3=-v}$.
% The above equation, Eq. \eqref{met2}, shows that the dimension of the parameter m and $v$ is $lenght^{-1}$ when $n=4$.
Equation (\ref{met2}) demonstrates that in the absence of electric charge ($v=0$), we revert to the non-charged case of TEGR, as indicated by Eq. (\ref{Tor1}) \cite{Nashed:2019zmy}. Moreover, it is a known fact that in GR and in the case of spherical symmetry, the $g_{tt}$ component is expected to be the inverse of the $g_{rr}$ component. However, this condition, as indicated by Eq. (\ref{met2}), is not satisfied unless $\mathit{v=0}$. The  asymptotic behavior of the spacetime (\ref{met2}) behaves as AdS/dS according to the sign of the cosmological constant $\Lambda$ \cite{Awad:2017tyz}.\vspace{0.2cm}\\

\underline{Singularity:}\vspace{0.2cm}\\
 To analyze the singularity, one needs to determine the values of  $r$ at which $a(r)$ and $a_1(r)$ become either a value of zero or infinity.  As singularities can be attributed to coordinate issues, the customary approach involves investigating different invariants that are constructed using both the Levi-Civita affine connection and the Weitzenb\"{o}ck connection. The scalars of the metric (\ref{met2}), derived from curvature using the Levi-Civita connection and from torsion using the Weitzenb\"{o}ck connection, are:\footnote{{ It is well-known that the curvature derived from the Levi-Civita connection can be expressed as:
 \begin{align}
 R^\mu{}_{\nu \rho \sigma}=\partial_\rho \overset{\mathbf{L}}{\Gamma}{}^\mu{}_{\nu\sigma}-\partial_\sigma \overset{\mathbf{L}}{\Gamma}{}^\mu{}_{\nu\rho}+\overset{\mathbf{L}}{\Gamma}{}^\mu{}_{\tau\rho} \overset{\mathbf{L}}{\Gamma}{}^\tau{}_{\nu\sigma}-\overset{\mathbf{L}}{\Gamma}{}^\mu{}_{\tau\sigma} \overset{\mathbf{L}}{\Gamma}{}^\tau{}_{\nu\rho}\,,\nonumber
 \end{align}
 where $\overset{\mathbf{L}}{\Gamma}{}^\mu{}_{\nu\sigma}$ is the Levi Civita affine connection defined as \[\overset{\mathbf{L}}{\Gamma}{}^\mu{}_{\nu\sigma}=g^{\mu \alpha}\left(g_{\alpha \nu,\sigma}+g_{\alpha\sigma, \nu}-g_{\sigma \nu,\alpha}\right)\,,\]  where  comma refers to partial derivative. Similarly, the curvature constructed from the  Weitzenb\"{o}ck  connection takes the form:
 \begin{align}
 R^\mu{}_{\nu \rho \sigma}=\partial_\rho \overset{\mathbf{w}}{\Gamma}{}^\mu{}_{\nu\sigma}-\partial_\sigma \overset{\mathbf{w}}{\Gamma}{}^\mu{}_{\nu\rho}+\overset{\mathbf{w}}{\Gamma}{}^\mu{}_{\tau\rho} \overset{\mathbf{w}}{\Gamma}{}^\tau{}_{\nu\sigma}-\overset{\mathbf{w}}{\Gamma}{}^\mu{}_{\tau\sigma} \overset{\mathbf{w}}{\Gamma}{}^\tau{}_{\nu\rho}\,.\nonumber
 \end{align}
The curvature constructed by the Weitzenb\"{o}ck connection vanishes completely, while the curvature formed by the Levi-Civita affine connection does not. \cite{Arcos:2004tzt}. }}
 \begin{eqnarray} \label{inv}
 && K=R^{\mu \nu \lambda \rho}R_{\mu \nu \lambda \rho} = \frac{f_1(r)}{(n-3)^6r^{2(n-6)}r^{6(n+2)}\Big\{1+\frac{(2n-5)v^{n-2}}{(n-3)r^{n-2}}+\frac{(2n-3)v{}^{n}}{(n-3)r^{n}}+\frac{(2n-1)v{}^{n+2}}{(n-3)r^{n+2}}\Big\}^{6}}\,,\nonu
&& R^{\mu \nu}R_{\mu \nu} =\frac{f_2(r)}{(n-3)^6r^{2(n-6)}r^{6(n+2)}\Big\{1+\frac{(2n-5)v^{n-2}}{(n-3)r^{n-2}}+\frac{(2n-3)v{}^{n}}{(n-3)r^{n}}+\frac{(2n-1)v{}^{n+2}}{(n-3)r^{n+2}}\Big\}^{6}}\,,\nonu
&&R =\frac{f_3(r)}{(n-3)^3r^{(n-6)}r^{3(n+2)}\Big\{1+\frac{(2n-5)v^{n-2}}{(n-3)r^{n-2}}+\frac{(2n-3)v{}^{n}}{(n-3)r^{n}}+\frac{(2n-1)v{}^{n+2}}{(n-3)r^{n+2}}\Big\}^{6}}\,,\nonu
&&T^{\mu \nu \lambda}T_{\mu \nu \lambda}= \frac{f_4(r)}{(n-3)^3r^{(n-6)}r^{2(n+2)}\Big\{1+\frac{(2n-5)v^{n-2}}{(n-3)r^{n-2}}+\frac{(2n-3)v{}^{n}}{(n-3)r^{n}}+\frac{(2n-1)v{}^{n+2}}{(n-3)r^{n+2}}\Big\}^{2}}\,, \nonu
&& T^\mu T_\mu =\frac{f_5(r)}{(n-3)^3r^{(n-6)}r^{2(n+2)}\Big\{1+\frac{(2n-5)v^{n-2}}{(n-3)r^{n-2}}+\frac{(2n-3)v{}^{n}}{(n-3)r^{n}}+\frac{(2n-1)v{}^{n+2}}{(n-3)r^{n+2}}\Big\}^{2}}\,, %\nonu
%
%&&T(r)=\frac{f_6(r)}{(n-3)^3r^{(n-6)}r^{2(n+2)}\Big\{1+\frac{(2n-5)v^{n-2}}{(n-3)r^{n-2}}+\frac{(2n-3)v{}^{n}}{(n-3)r^{n}}+\frac{(2n-1)v{}^{n+2}}{(n-3)r^{n+2}}\Big\}^{2}},
\end{eqnarray}
where $f_i(r), i=1\cdots 5$ are polynomial functions of $r$.    { Eq.~\eqref{inv}, demonstrates that all invariants derived from curvature and torsion exhibits a genuine singularity located at $r = 0$}. As we approach the limit of the aforementioned invariants with the value of $r$ approaching zero, we obtain  $K= R^{\mu \nu}R_{\mu \nu}\sim r^{-2(n-6)}$, $R=T^{\mu \nu \lambda}T_{\mu \nu \lambda}=T^{\mu}T_{\mu}\sim r^{-(n-6)}$. As $r\to \infty$ we get  $ R^{\mu \nu}R_{\mu \nu}=R=T^{\mu \nu \lambda}T_{\mu \nu \lambda}=T^{\mu}T_{\mu}\sim r^{-(n-2)}$,
which contradicts the solution proposed by Einstein-Maxwell theory in the framework of TEGR  theory that has $K= R^{\mu \nu}R_{\mu \nu}\sim r^{-2n}$ and $R=T^{\mu \nu \lambda}T_{\mu \nu \lambda}=T^{\mu}T_{\mu}\sim r^{-n}$ respectively. This clearly shows that there is  a much softer singularity than in the TEGR case  and softer singularity than in the  Awad et al solution \cite{Awad:2017tyz}. It is imperative to mention that the softness of the singularities is due to the contribution of the higher order charged terms. There is now a question ``according to Tipler and Krolak \cite{Tipler:1977,CLARKE1985127} are these singularities weak? If so, can we extend the geodesics behind these regions? These questions need more study, which will be done elsewhere.  \vspace{0.1cm}\\

 Using Eq. (\ref{Tor1}), after writing $r$ in terms of $T$ asymptotically,  in (\ref{sol}) we get
 \be \label{fT} f(T)\simeq 2\Lambda+\frac{T}{44\Lambda}-\frac{T^2}{792\Lambda^2}+O\left(T^3\right)+\cdots.\ee  { As Eq. (\ref{fT}) shows that when the scalar torsion, $T$ is vanishing or equal constant  we get a constant value of  $f(T)$. If we compare this result with Eq. (\ref{sol}) we see that $f(r)$ is constant when the dimensional constant $c_3=0$ and this makes the scalar torsion $T$ has a constant values as Eq. (\ref{Tor1}) shows. This means that for non-constant torsion scalar and in the frame of cylindrical coordinate we should include the electric field. In this special case we can get a non trivial form of $f(T)$ as Eq. (\ref{fT}) shows. }. The behavior of $f(T)$ is drawn in figure \ref{Fig:2}\subref{fig:2b}.

  Subsequently, we will move forward with the calculation of the energy associated with the obtained black hole solutions, as described by Eq. (\ref{sol})\footnote{Within this inquiry, we treat the Newtonian constant as an effective coefficient designated by $G_{\text{eff.}}$, where $G_{\text{eff.}}=\frac{G_{N}}{f_T}$, as discussed in \cite{Wei:2011mq}.}. Using the definition of the supperpotential tensor $S^{\mu \nu \rho}$, we evaluate the essential energy components of the solution given by Eq. (\ref{sol}) as follows:
\be \label{sup} S^{001}=\frac{(n-2)}{2(n-3)r\Big\{1+\frac{(2n-5)v^{n-2}}{(n-3)r^{n-2}}+\frac{(2n-3)v{}^{n}}{(n-3)r^{n}}+\frac{(2n-1)v{}^{n+2}}{(n-3)r^{n+2}}\Big\}^{2}}.  \ee
%\be S^{001}=\frac{c_2{}^2c_3-4c_4c_2{}^2r^3-6c_1{}^2 r^3}{144r^4\beta c_4(c_1{}^2-c_2{}^2)^2},  \ee
 Upon inserting Eq. (\ref{sup}) into Eq. (\ref{q3}), the energy takes the subsequent expression:
{\ba \label{e1} &&P^0=E= -\frac{\Omega_{{n}-1}}{\kappa}\Bigg(2(n-2)\Lambda r^{(n-1)}-\frac{(2n-5)(n-2)m^{n-3}}{n-1}-\frac{(2n-7)v^{n-3}}{(n-3)r^{n-3}}-\frac{(n-2)v^{2n-5}}{(n-1)r^{2n-5}}
-\frac{2(n-2)v^6}{(3n-4)r^{2n-3}}\nonumber\\
&&-\frac{2(n-2)v^8}{(3n-2)r^{2n-1}}\Big).\ea}
Equation (\ref{e1}) shows that the value  of energy is divergent thus  we should use the regularized method  to get finite value. Using Eq. (\ref{q3}) we get the regularized form of the energy in the shape
\be \label{q444} P^a:=\frac{1}{\kappa}\int_V d^{n-2}x  \left[e{S}^{a 0
0} f_T\right]-\frac{1}{\kappa}\int_V d^{n-2}x \left[e{S}^{a 0
0} f_T\right]_{\it the\; physical\;  quantities\; equal\; zero}.\ee In this context, ``the physical quantities equal zero'' refers to the condition when the constants $m=0$.  Using (\ref{q444}) in solution (\ref{sol}) we get:
{\ba \label{e2} P^0=E&=& \frac{\Omega_{{n}-1}}{\kappa}\Bigg[\frac{(2n-5)(n-2)m^{n-3}}{n-1}+\frac{(2n-7)v^{n-3}}{(n-3)r^{n-3}}+\frac{(n-2)v^{2n-5}}{(n-1)r^{2n-5}}+\frac{2(n-2)v^{2n-3}}
{(3n-4)r^{2n-3}}+\frac{2(n-2)v^{2n-1}}{(3n-2)r^{2n-1}}
\Bigg].\ea}
{ Equation (\ref{e2}) shows that we have the ADM mass when the dimensional constant $v=0$ and $n=4$ otherwise, we have contributions from the charged terms. Equation (\ref{e2}) is consistent with what was obtained before \cite{Awad:2017tyz} up to the cubic term of the charge \cite{Nashed:2003ee,Nashed:2011fg}.}
\section{Black hole rotation in the context of nonlinear Maxwell-$\mathit{f(T)}$ }\label{S5}
Next, we will work towards acquiring a solution for a rotating black hole that adheres to the field equations outlined in Eq. (\ref{q2}) within the framework of $\mathit{f({T})}$ gravity. Our approach will involve constraining our analysis to the static black hole solution (\ref{sol}), followed by the application of the subsequent transformation:
\begin{equation} \label{t1}
\bar{\psi}_{i} =-\Xi~ {\psi_{i}}+\frac{ \phi_i}{l^2}~t,\qquad \qquad \qquad
\bar{t}=
\Xi~ t-\sum\limits_{i=1}^{\omega}\phi_i~ \psi_i,
\end{equation}
where $\phi_i$ represents the rotation parameters, and their count is denoted as $\omega = \lfloor(n - 1)/2\rfloor$. Moreover, the parameter $l$ is linked to the cosmological constant $\Lambda$  as follows:
\begin{eqnarray}
l=-\frac{(n-2)(n-1)}{2 \Lambda}.
\end{eqnarray}
Moreover, $\Omega$ is figured as:
\[\Omega:=\sqrt{1-\sum\limits_{j=1}^{{\omega}}\frac{\phi_j{}^2}{l^2}}.\]
By implementing the transformation (\ref{t1}) on the vielbein (\ref{tetrad}), we acquire
\begin{eqnarray}\label{tetrad1}
\nonumber \left({b^{i}}_{\mu}\right)=\left(
  \begin{array}{cccccccccccccc}
    \Omega\sqrt{a(r)} & 0 &  -\phi_1\sqrt{a(r)}&-\phi_2\sqrt{a(r)}\cdots &
-\phi_{\omega}\sqrt{a(r)}&0&0&\cdots&0
\\[5pt]
    0&\frac{1}{\sqrt{a_1(r)}} &0 &0\cdots &0&0&0&\cdots & 0\\[5pt]
          \frac{\phi_1r}{l^2} &0 &-\Omega r&0  \cdots &0&0&0&\cdots & 0\\[5pt]
        \frac{\phi_2r}{l^2} &0 &0  &-\Omega r\cdots & 0&0&0&\cdots & 0\\[5pt]
        \vdots & \vdots  &\vdots&\vdots&\vdots &\vdots&\vdots& \cdots & \vdots \\[5pt]
 \frac{ \phi_\omega r}{l^2}  &  0 &0&0 \cdots & -\Omega r&0&0&\cdots & 0 \\[5pt]
   0 &  0  &0&0 \cdots &0&r&0&\cdots & 0\\[5pt]
     0 &  0  &0&0 \cdots &0&0&r&\cdots & 0\\[5pt]
       0 &  0 &0&0 \cdots &0&0&0&\cdots & r\\
  \end{array}\right)\,.&\\
\end{eqnarray}
  Incorporating the static solution (\ref{sol}) for the functions $a(r)$ and $a_1(r)$ derived earlier, the electromagnetic potential $A(r)$ assumes the following form following the application of transformation (\ref{t1}):
\begin{equation}
\label{Rotpot}
\bar{A}(r)=-A(r)\left[\sum\limits_{j=1}^{n} \phi_j d\psi'_j-\Omega dt'\right].
\end{equation}
It's important to observe that the two sets of vielbeins, as described in equations (\ref{tetrad}) and (\ref{tetrad1}), can be locally transformed into one another, but this equivalence does not hold globally \cite{Lemos:1994xp, Awad:2002cz}. The metric corresponding to the vielbein (\ref{tetrad1}) is expressed as follows:
\begin{eqnarray}
\label{m1}
    ds^2=-a(r)\left[\Omega d{\bar {t}}  -\sum\limits_{i=1}^{\omega}  \phi_{i}d{\bar {\psi_1}}
\right]^2+\frac{dr^2}{a_1(r)}+\frac{r^2}{l^2}\sum\limits_{i=1 }^{\omega}\left[\phi_{i}d{\bar {t}}-\Omega l^2 d{\bar{\psi}}_i\right]^2+
r^2 d\xi_k^2-\frac{r^2}{l^2}\sum\limits_{i<j
}^{\omega}\left(\phi_{i}d{\bar {\psi}}_j-\phi_{j}d{\bar {\psi}}_i\right)^2\,.
\end{eqnarray}
In this context, the variables have the following ranges: $0\leq r< \infty$, $-\infty < t < \infty$, $0 \leq \Omega_{i}< 2\pi$, where $i=1,2 \cdots \omega$, and $-\infty < \xi_k < \infty$ for $k = 1, 2, \ldots, n-3$. The expression $d \xi_k^2$ denotes the Euclidean metric in a space with $(n-\omega-2)$ dimensions.

It's crucial to emphasize that Eq.~ (\ref{met2})  can be obtained as a particular instance of the broader metric described earlier by assigning a value of zero to all rotation parameters $\phi_j$. Additionally, it's worth noting that Eq.~(\ref{m1}) is obtained when Eq.~ (\ref{met1})  is applied as elaborated in \cite{Nashed:2018cth}.
\begin{eqnarray}
\label{min}
\nonumber \left(\eta_{ij}\right)=\left(
  \begin{array}{cccccccccccccc}
    -1 & 0 & 0&0&0&0&0&0&0&\cdots&0 \\[5pt]
    0&1 &0 &0 &0&0&0&0&0&\cdots & 0\\[5pt]
        0 &0 &1+\frac{\phi_\omega{}^2}{l^2\Omega^2} &-\frac{\phi_\omega \phi_1}{l^2\Omega^2}
&-\frac{\phi_\omega \phi_2}{l^2\Omega^2}&\cdots &-\frac{\phi_\omega \phi_{\omega-1}}{l^2\Omega^2}&0&0&\cdots &0\\[5pt]
        0 &0 &-\frac{\phi_\omega \phi_1}{l^2\Omega^2}
&1+\frac{\phi_{\omega-1}{}^2}{l^2\Omega^2} & -\frac{\phi_{\omega-1} \phi_1}{ l^2\Omega^2}&\cdots&-\frac{\phi_{\omega-1} \phi_{\omega-2}}{l^2\Omega^2}
&0&0&\cdots & 0\\[
5pt]
        \vdots & \vdots  &\vdots&\vdots&\vdots &\vdots&\vdots& \cdots & \vdots \\[5pt]
  0 &0 &-\frac{\phi_\omega \phi_{\omega-1}}{l^2\Omega^2} &-\frac{\phi_{\omega-1}\phi_{\omega-2}}{l^2\Omega^2}
&-\frac{\phi_{
\omega-2}\phi_{\omega-3}}{l^2\Omega^2}&\cdots &1+\frac{\phi_1{}^2}{l^2\Omega^2}&0&0&\cdots
&0\\[5pt]
   0 &  0  &0&0  &0&\cdots&0&1&0&\cdots & 0\\[5pt]
     0 &  0  &0&0 &0&\cdots&0&0&1&\cdots & 0\\[5pt]
       0 &  0 &0&0 &0&\cdots&0&0&0&\cdots & 1\\
 \end{array}
\right).&\\
\end{eqnarray}

 %%%%%%%%%%%%%%%%%%%%%%%%%%%% Section 7 %%%%%%%%%%%%%%%%%%%%%%%%%%%%%
\section{Thermodynamic properties of  solution (\ref{sol})}\label{S6}
%%%%%%%%%%%%%%%%%%%%%%%%%%%%%%%%%%%%%%%%%%%%%%%%%%%%%%%%%%%%%%%%%%%%
 The definition of the Hawking temperature is as follows \cite{Sheykhi:2012zz,Sheykhi:2010zz,Hendi:2010gq,Sheykhi:2009pf}:
  \begin{equation}
T_+ = \frac{a'(r_+)}{4\pi}\,.
\end{equation}
The  horizon is existed  at $r = r_+$, and it meets the condition where $a'(r_+) \neq 0$. In the context of $\mathit{f({T})}$ theory, the entropy is expressed as per \cite{Cai:2015emx}:
\begin{equation}\label{ent}
S(r_+)=\frac{{\cal A}}{4}\frac{df(T)}{dT}=\frac{{\cal A}}{4}\frac{df(r)}{dr}\frac{dr}{dT(r)}.
\end{equation}
In Equation (\ref{ent}), the entropy is symbolized by the term involving ${\cal A}$, which signifies the area. It's important to highlight that, unlike in General Relativity (GR), the entropy is not directly proportional to the area. This is due to the additional influence of the first derivative of $f(T)$. However, when $f(T)=T$, corresponding to the TEGR scenario, the entropy reverts to the one found in GR.

The stability of the solution hinges on the heat capacity's sign, denoted as $C$. Specifically, if $C>0$, the model is considered stable; otherwise, it is deemed unstable. Our next step involves examining the  stability of the solution by analyzing the behavior of its heat capacities, as defined in references \cite{Nouicer:2007pu, DK11, Chamblin:1999tk}:
\begin{equation}\label{m55}
C_+=\frac{\partial m}{\partial r_+} \left(\frac{\partial T}{\partial r_+}\right)^{-1}.
\end{equation}
%Therefore, if $(C_{+} > 0, C_+ < 0)$, then the black hole is thermodynamically stable or unstable.

\begin{figure}
\centering
\subfigure[~ $f(r)$ given by (\ref{sol})]{\label{fig:3a}\includegraphics[scale=0.35]{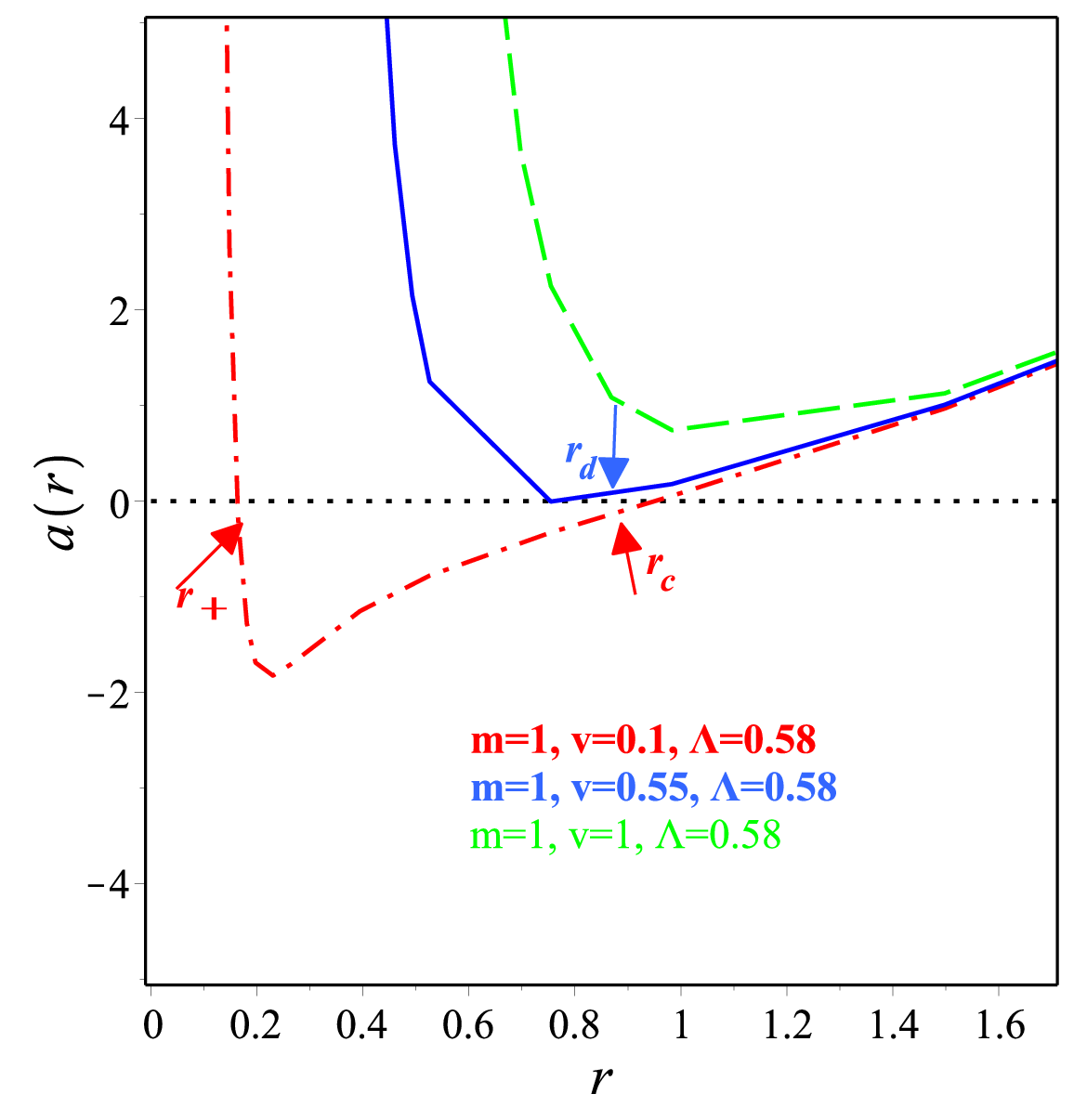}}
\subfigure[~The analytic function $f(T)$ given by (\ref{fT})]{\label{fig:3b}\includegraphics[scale=0.35]{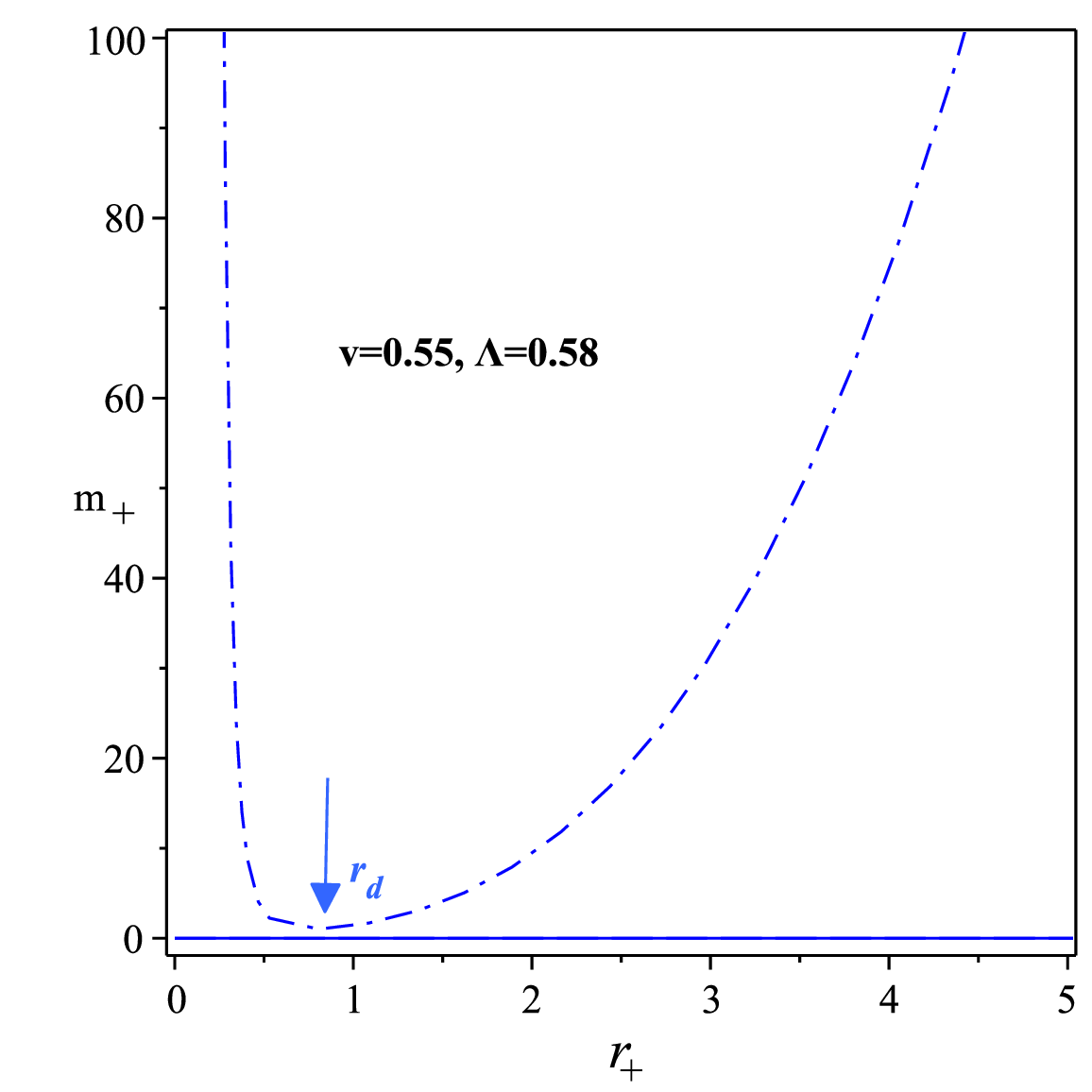}}
\caption{Schematic plot of the functions $a(r)$ and $m_+$.}
\label{Fig:3}
\end{figure}

In our investigation of the various thermodynamic properties using the black hole solution (\ref{sol}), we initiate the analysis with the condition $a(r_+) = 0$ (please note that we are restricting our study to a four-dimensional spacetime for the examination of thermodynamic quantities). This condition leads to the following expression:
\begin{eqnarray} \label{m33}
&& m_+=\frac{v{}^2}{2r_+}+\frac{v{}^4}{3r_+{}^3}+\frac{v{}^6}{4r_+{}^5}
+\frac{v^8}{5r_+{}^7}
+\Lambda r_+{}^3\,.\end{eqnarray}
 Equation (\ref{m33}) reveals that $M$ of the  solution (\ref{sol}) is dependent on $r$  of the horizons. Figure \ref{Fig:3}\subref{fig:3a} displays the relationship between the   $a(r)$ and   $r$ for a 4-dimensional case, illustrating the potential horizons of the solution. Furthermore, Figure \ref{Fig:3}\subref{fig:3b} illustrates the correlation between $M$ and $r$.

Utilizing Eq. (\ref{ent}), we evaluate the entropy of solution (\ref{sol}) in the following manner:
\begin{eqnarray} \label{ent1}
{S_+}=4\pi\Big[ r_+{}^2+3v^2+\frac{5v^4}{r_+{}^2}+\frac{7v^6}{r_+{}^4}\Big].
\end{eqnarray}
From Eq. (\ref{ent1}), we can deduce the following observations:\vspace{0.2cm}\

\begin{figure}
\centering
\subfigure[~Entropy of solution (\ref{sol})]{\label{fig:4a}\includegraphics[scale=0.3]{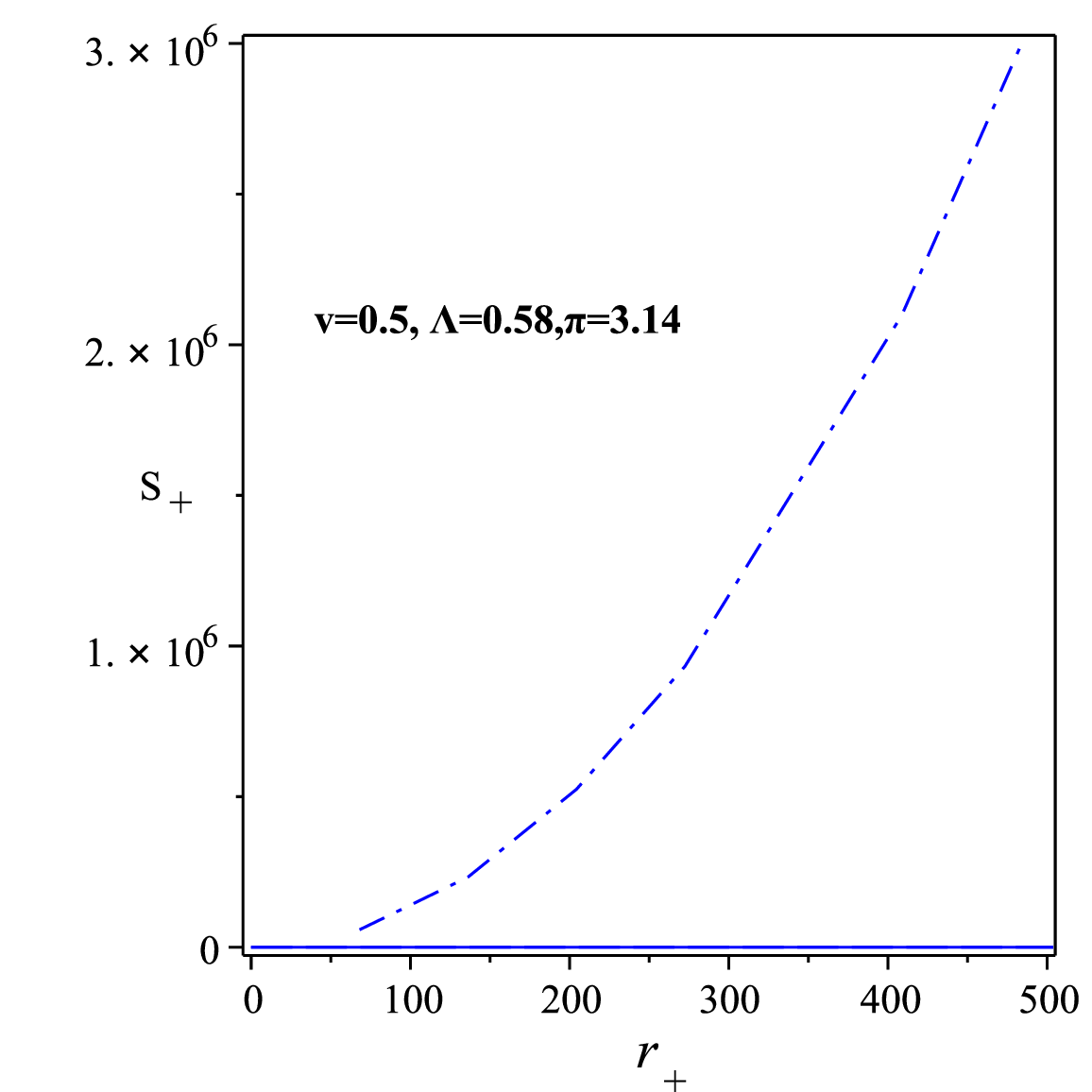}}
\subfigure[~Temperature of  solution (\ref{sol})]{\label{fig:4b}\includegraphics[scale=0.3]{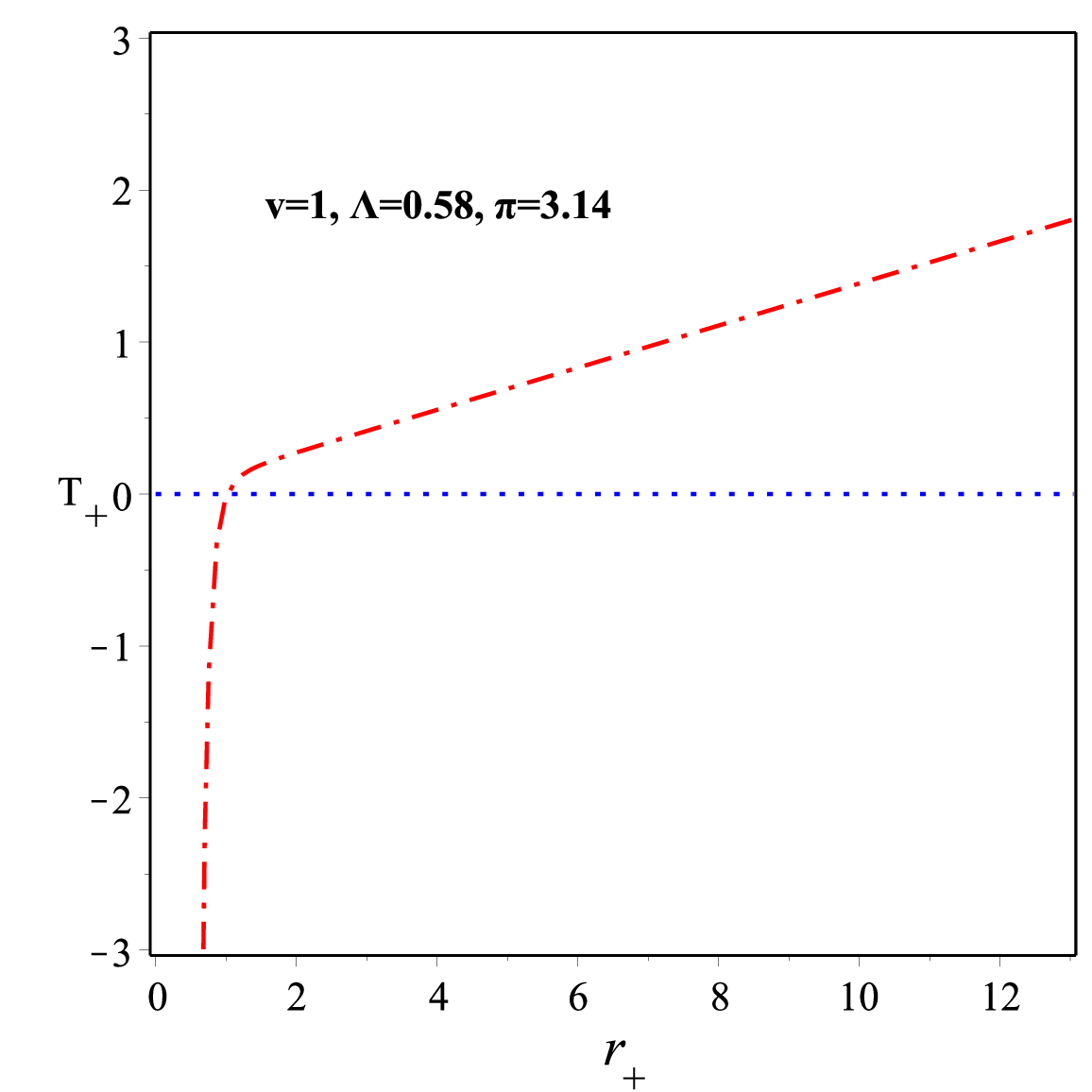}}
\subfigure[~Heat capacity of  solution (\ref{sol})]{\label{fig:4c}\includegraphics[scale=0.3]{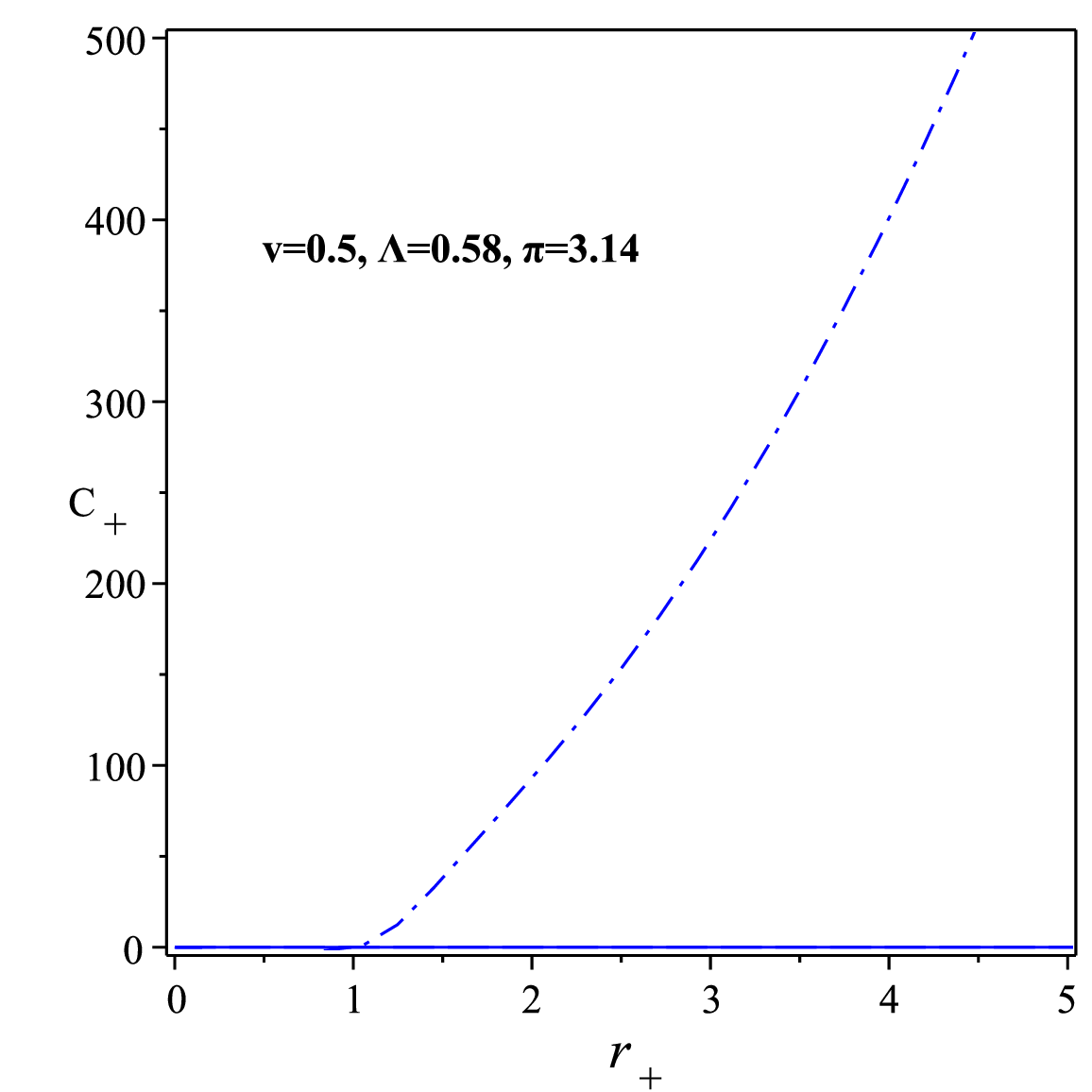}}
\caption{Schematic plot of the entropy and temperature.}
\label{Fig:4}
\end{figure}

The entropy doesn't exhibit a straightforward proportionality to the area, primarily due to the significant contributions stemming from the electric charge.\vspace{0.2cm}\
When $v=0$ and $n=4$, Eq. (\ref{ent1}) coincides with that of TEGR theory, implying that $S_+$ becomes directly proportional to the area ($A$).\vspace{0.2cm}\
Figure \ref{Fig:4}\subref{fig:4a} depicts the trend of entropy in four dimensions, revealing a positive magnitude.

We determine the temperature linked to   solution (\ref{sol}) in the following manner:
\begin{eqnarray} \label{m44}
{T_+}=\frac{1}{4\pi}\Big[3\Lambda r_+-\frac{v^2}{4r_+{}^3}-\frac{v^4}{2r_+{}^5}-\frac{9v^6}{8r_+{}^7}-\frac{7v^8}{10r_+{}^{9}}\Big]\,.\end{eqnarray}
The Hawking temperature at the event horizon, denoted as ${T_+}$, is depicted in Figure \ref{Fig:4}\subref{fig:4b}. The plot illustrates a transition from a negative value to a positive value. The presence of this negative Hawking temperature is notable, as it signifies the formation of an ultracold black hole. This discovery is consistent with Davies' work \cite{Davies:1978mf}, which demonstrated that there is no inherent contradiction preventing a black hole's temperature from being negative, leading it to transition into a naked singularity. As Figure \ref{Fig:4}\subref{fig:4b} demonstrates, this scenario is indeed observed. The occurrence of an ultracold black hole can be attributed to the existence of a phantom energy field \cite{Babichev:2014lda}.

We will now proceed to evaluate the heat capacity of Eq.(\ref{sol}). Employing Eq. (\ref{m55}), we derive the following expression:
\begin{equation} \label{enr}
C(r_+)=\frac{4\pi \Big[3\Lambda r_+{}^{2}-\frac{v^2}{2r_+{}^{2}}-\frac{v^4}{r_+{}^{4}}-\frac{5v^6}{4r_+{}^{6}}-
\frac{7v^8}{5r_+{}^{8}}\Big]}{3\Lambda +\frac{3v^2}{4r_+{}^{4}}+\frac{5v^4}{2r_+{}^{6}}+\frac{63v^6}{8r_+{}^{8}}+\frac{63v^8}{10r_+{}^{10}}}.
\end{equation}
%\begin{figure}
%\centering
%\includegraphics[scale=0.4]{JRf}
%\includegraphics[scale=0.40]{JRHeat}
%\caption{Schematic plot of the heat capacity of the black hole (\ref{sol}).}
%\label{Fig:5}
%\end{figure}
 We have plotted the heat capacity in Figure \ref{Fig:4}\subref{fig:4c}, which shows a positive value. The thermodynamic stability of charged black hole spacetime has been extensively studied in various theories. This includes thermodynamics of Bardeen (regular) black holes \cite{Myung:2007qt}, Schwarzschild-AdS in two vacuum scales case \cite{Dymnikova:2010zz}, and similar works in noncommutative geometry \cite{Berej:2006cc,Man:2013hpa,Tharanath:2014naa,Maluf:2018lyu}.

%The free energy in the grand canonical ensemble which  called Gibb's free energy is defined as \cite{Zheng:2018fyn,Kim:2012cma}
%\begin{equation} \label{enr}
%G(r_h)=M(r_+)-T(r_+)S(r_+)%+P(r_+)V(r_+),
%\end{equation}
%$V$ is the geometric volume  of the black hole, $P$ is the pressure which is represented by the radial    equation of (\ref{f1}), i.e. $I_r{}^r$,
%with    $T(r_h)$,  $S(r_h)$ and $E(r_h)$ are  the temperature, the entropy and  the quasilocal energy at the event horizon.   Using Eqs. (\ref{ent}), %(\ref{m33}), (\ref{ent1}) and (\ref{m44}) in (\ref{enr}) we get
%\begin{eqnarray} \label{m77}
%&&{G_+}=r^{d-1}\Lambda-\frac{1}{2(d-2)}c_3{}^2r^{{d-2}/2}\sqrt{d^{d-3}+c_2}.
%\end{eqnarray}
%\begin{figure}[ht]
%\centering
%\includegraphics[scale=0.4]{JFT5.eps}
%\caption{{\it{The Gibb's free energy.}}
% }
%\label{Fig:5}
%\end{figure}

%where we have put $V(r_+)=\frac{4}{3}\pi r_+{}^3$ and
%The behaviors of the Gibb's energy of the black holes (\ref{sol}) is presented in figure  \ref{Fig:8}  for particular values of the model parameters. As figure \ref{Fig:8}  shows that, for the black hole solution (\ref{sol}), Gibb's energy is always positive when $\alpha>0$   which means that it is more
%globally stable.

 \section{Discussion}\label{S7}
The purpose of this research was twofold:  The initial objective of this study was to obtain a general expression for $f(T)$ by considering a cylindrical spacetime in $n$ dimensions. To aim this we applied the charged field equation to n-dimensional { cylindrical}  spacetime having two unknown functions. The resulting non-linear differential equations are solved using the chain rule by assuming $f(T)=f(r)$, since we study spherical spacetime, $f_T=\frac{df(T)}{dT}=\frac{df(r)}{dr}\frac{dr}{dT}$ and  $f_{TT}=\frac{d^2f(T)}{dT^2}=\frac{dfT}{dT}=\frac{d}{dr}\Big[\frac{df(r)}{dr}\frac{dr}{dT}\Big]\frac{dr}{dT}$. Using this procedure we succeeded to derive an analytic static solution. One intriguing characteristic of this black hole is that it exhibits a non-constant value of the torsion scalar, denoted as $T(r)$. From the value of the torsion scalar, one can write the radial coordinate in terms of the torsion scalar, i.e., $r(T)$. By incorporating this value in the form of $f(r)$, one can determine the specific expression for $f(T)$.  From the above discussion, we have the following comments:\vspace{0.2cm}\\
1- The form of the two unknown functions that are presented in the vielbein have two dimensional constants one related to the mass and the other to the electric charge (both have unit of length). The substantial property is the fact that the constant that linked to the electric charge has higher order contributions and if this constant is set equal zero we get a solution of the TEGR case and in that situation, the function $f(T)$ has a constant value. \vspace{0.2cm}\\
2- In spite that we have a spherical spacetime the two-components of the metric potential $g_{tt}$ and $g_{rr}$ are not proportional to each other, i.e., $g_{tt} {g_{rr}}{}^{-1}\neq 1$ however, their Killing and event horizons are equal.

The second aim of this research was to study the physical properties of the derived solution.To achieve this objective, we examine the invariants formed using curvature and torsion. %One more interesting property of this black hole is the fact that it possesses related to the singularity.
 We have demonstrated that  solution remains devoid of singularities when $4\leq n\leq 7$. In the case of $n>7$, we encounter a significantly milder singularity compared to the scenario in TEGR. Furthermore, it represents a softer singularity in contrast to the black hole solution derived in \cite{Awad:2017tyz}. The main reason for the weakness of the singularity comes form the contributions of the higher order terms of the charge. Moreover, we calculated the energy, using the conserved  (n-1)-dimensional vector momentum,  of this solution and shown that it is coincides with the ADM mass up to order $\Big(\frac{1}{r}\Big)$ and if we take into account the higher  of $\Big(\frac{1}{r}\Big)$ we got a contributions from the higher order charged terms. Also we applied a coordinate transformation to the black hole solution and derived a stationary rotating black hole solution having $\omega$ rotation parameters with non trivial value of $T$ and  $f(T)$.

  Finally, we have scrutinized the thermal stability of the static solution. In pursuit of this objective, we computed the Hawking temperature and observed that it descends below absolute zero, resulting in the formation of an ultracold black hole.  Furthermore, we calculated the heat capacity and demonstrated that it always remains positive, indicating that the static black hole possesses thermodynamic stability. {\it It is essential to stress on the fact that the procedure applied in this study is not successful to apply to the spherically symmetric spacetime studied in \cite{2012MNRAS.427.1555I}. The main reason for this failure comes from the appearance of the $r, \theta$ component in the field equations which restricts the solution to be coincides with the TEGR case.
 %%%%%%%%%%%%%%%%%%%%%%%%%%%%%%%%%%%%%%%%%%%%%%%%%%%%%%%%%%%%%%%%%%%%%%%%%%%%
%\subsection*{Acknowledgments}
%We would like to thank S. Capozziello and E. N. Saridakis for useful discussions.
 %This work is partially supported by the Egyptian Ministry of Scientific Research under project No. 24-2-12.
%%%%%%%%%%%%%%%%%%%%%%%%%%%%%%%%%%%%%%%%%%%%%%%%%%%%%%%%%%%%%%%%%%%%%%%%%%%%%%%%%%%%%%
%\bibliographystyle{apsrev}
%\bibliography{Ref}
%%%%%%%%%%%%%%%%%%%%%%%%%%%%%%%%%%%%%%%%%%%%%%%%%%%%%%%%%%%%%%%%%%%%%%%%%%%%%%%%%%%%%%

\end{document}